\documentclass[11pt,a4paper]{article}
\pdfoutput=1

\usepackage{mathtools}
\usepackage{mathrsfs}
\usepackage{amsmath}
\usepackage[utf8x]{inputenc}
\usepackage{amssymb}
\usepackage{slashed}
\usepackage{bm}
\usepackage{color}
\usepackage{tikz}
\usepackage{cite}
\usepackage{braket}
\usepackage{float}
\usepackage[]{todonotes}
\usepackage{jheppub}
\usepackage{twistor}

\makeatletter
\renewcommand{\todo}[2][]{%
    \@todo[caption={#2}, #1]{\begin{spacing}{0.5}#2\end{spacing}}%
} 
\makeatother 
\usepackage{setspace}

\newcommand{\ud}{{\mathrm{d}}}
\hfuzz=\maxdimen \tolerance=10000
\vfuzz=\maxdimen \tolerance=10000
\newcommand{\LCm}{{\scriptscriptstyle -}}
\newcommand{\LCp}{{\scriptscriptstyle +}}

\newcommand{\LCperp}{{\scriptscriptstyle \perp}}

\title{All order gravitational waveforms from scattering amplitudes}
\author[1]{Tim Adamo,}
\author[1]{Andrea Cristofoli,}
\author[2]{Anton Ilderton}
\author[1]{and Sonja Klisch}

\affiliation[1]{School of Mathematics and Maxwell Institute for Mathematical Sciences\\
University of Edinburgh, EH9 3FD, UK}
\affiliation[2]{Higgs Centre, School of Physics and Astronomy\\ University of Edinburgh, EH9 3FD, UK}

\emailAdd{t.adamo@ed.ac.uk}
\emailAdd{acristof@exseed.ed.ac.uk}
\emailAdd{anton.ilderton@ed.ac.uk}
\emailAdd{s.klisch@ed.ac.uk}

\abstract{Waveforms are classical observables associated with any radiative physical process.  Using scattering amplitudes, these are usually computed in a weak-field regime to some finite order in the post-Newtonian or post-Minkowskian approximation. Here, we use strong field amplitudes to compute the waveform produced in scattering of massive particles on gravitational plane waves, treated as exact nonlinear solutions of the vacuum Einstein equations. Notably, the waveform contains an infinite number of post-Minkowskian contributions, as well as tail effects. We also provide, and contrast with, analogous results in electromagnetism.}

\notoc
\begin{document}

\maketitle


\clearpage
\pagenumbering{arabic}

\section{Introduction}

The observation of gravitational waves has brought renewed importance to the study of general relativity and its observables. Surprisingly, scattering amplitudes -- one of the key outputs of \emph{quantum} field theory -- are providing a new way to study \emph{classical} general relativity; for reviews see~\cite{Bjerrum-Bohr:2022blt,Kosower:2022yvp,Buonanno:2022pgc}. Starting from novel perspectives on~\cite{Neill:2013wsa,Damour:2016gwp,Cheung:2018wkq}, and a remarkable state-of-the-art calculation for~\cite{Bern:2019nnu}, the conservative Hamiltonian of the gravitational two-body problem, a new program for providing higher-order post-Minkowskian (PM) approximations to gravitational observables has emerged based on the classical limit of scattering amplitudes. This has led to a variety of exciting new results for gravitational observables, e.g.~\cite{DiVecchia:2021ndb,Bern:2021dqo,Jakobsen:2021smu,DiVecchia:2021bdo,Bjerrum-Bohr:2021din,Jakobsen:2021lvp,Damgaard:2021ipf,Brandhuber:2021eyq,Monteiro:2021ztt,Bern:2021yeh,Jakobsen:2022fcj,Manohar:2022dea,Aoude:2022trd,DiVecchia:2022nna,Chen:2022yxw,Menezes:2022tcs,Jakobsen:2022psy,Kalin:2022hph,Damgaard:2022jem}, which build on many of the powerful structures in scattering amplitudes such as generalized unitarity and double copy, as well as techniques from effective field theory.

A key tool in this program has been the development of a formalism to systematise the extraction of classical physical observables from scattering amplitudes~\cite{Kosower:2018adc}. So far, all observables computed with this approach are valid for weak fields only: they are obtained from amplitudes at finite PM order, so truncate at a corresponding fixed order in the coupling~\cite{Maybee:2019jus,Guevara:2019fsj,Aoude:2021oqj,delaCruz:2020bbn,Cristofoli:2021vyo,Cristofoli:2021jas,Herrmann:2021lqe,Herrmann:2021tct,Britto:2021pud, Bautista:2021inx,Bautista:2021llr,Bautista:2019tdr}. This is in sharp contrast with other approaches to gravitational dynamics such as the self-force paradigm~\cite{Cutler:1994pb,Barack:2009ux,Poisson:2011nh,Harte:2018iim,Barack:2018yvs,Barack:2022pde}, where perturbation theory is implemented around a curved background and the weak field limit is not considered.

To address this gap, the amplitudes-based approach can be generalised to curved backgrounds by means of strong field scattering amplitudes and their classical limits~\cite{Adamo:2022rmp}. This provides an alternative route to the computation of classical observables, as strong field amplitudes encode a substantial amount of information about higher-order processes~\cite{tHooft:1987vrq,Amati:1987wq,Jackiw:1991ck,Kabat:1992tb,Giddings:2004xy,Lodone:2009qe,Adamo:2021hno,Adamo:2021rfq} and finite size effects~\cite{Bonnor:1969,Balasin:1994tb,Adamo:2022rob} in trivial backgrounds, and can also admit remarkably compact formulae~\cite{Adamo:2020syc,Adamo:2020yzi,Adamo:2022mev}. A key aspect is that even first order perturbation theory around a curved background -- which we refer to as `first post-background', or 1PB, order -- encodes \emph{infinitely} many orders of the PM expansion.
This is analogous to the relation between the PM and post-Newtonian (PN) expansions for bound orbits, where a fixed contribution of the former encodes infinitely many orders of the latter due to the virial theorem.  

Here we show for the first time how classical observables encoding all-order results can be extracted from scattering amplitudes. We derive expressions for the {classical} gravitational waveform emitted by a point particle scattering on a gravitational plane wave (an exact solution to the nonlinear Einstein equations), encoding all-order contributions in the PM expansion when the flat spacetime limit is taken, as well as tail effects which usually enter at high order in the PM approximation. {We also perform analogous calculations for charged particles scattering on electromagnetic plane waves. While our aim is not to study the phenomenology of electrodynamics, the waveforms do not seem to appear in an otherwise extensive literature~\cite{Ritus1985,DiPiazza:2011tq,Gonoskov:2021hwf,Fedotov:2022ely}, and it is revealing to compare and contrast with the gravitational case~\cite{OuldElHadj:2021fqi,Saketh:2021sri,Bern:2021xze,Audagnotto:2022lft}.} 

Note that plane waves are not just good models of gravitational waves, but also describe \emph{any} spacetime in the neighbourhood of a null geodesic~\cite{Penrose:1976}. This directly connects our results to the gravitational 2-body problem: in the limit where one mass is negligible, the massless probe will experience the heavy body's metric as a plane wave. Indeed, plane wave/ultrarelativistic limits have been used to analyse gravitational self-force~\cite{Galley:2013eba} and black hole quasinormal modes~\cite{Fransen:2023eqj}.

%
\section{Asymptotic waveforms}
Let $\ket{\Psi}$ be a normalised superposition of free particle (mass $m$) states,
\be
    \ket{\Psi} = \int\!\ud\Phi(p)\, \phi(p)\, \e^{i p \cdot b/\hbar}\, \ket{p} \;,
\ee
{where $\ud\Phi(p)$ is} the Lorentz-invariant on-shell measure, the wavepacket $\phi(p)$ has a well-defined classical limit (cf., \cite{Kosower:2018adc}) {and $b_\mu$ is the impact parameter}. This state is evolved on an electromagnetic or gravitational plane wave background.
In terms of the S-matrix $\cal S$ on that background, the time-evolved state is simply~${\cal S}\ket{\Psi}$. 

{Our interest is in the classical gravitational or  electromagnetic radiation emitted by a scalar particle as it scatters on these backgrounds, as measured by an asymptotic observer at future null infinity. The particular observable of interest is the waveform, encoded in the expectation value of the Maxwell and Riemann tensors, $\langle F_{\mu\nu}(x)\rangle$ and $\langle R_{\mu\nu\sigma\rho}(x)\rangle$. In coordinates $x^{\mu}=(t,\mathbf{x})$, approaching future null infinity corresponds to taking $r\equiv |{\bf x}|\to\infty$ while $u=t - r$ is held constant. Following~\cite{Cristofoli:2021vyo}, the waveform $W$ is defined simply as the coefficient of the leading $1/r$ term in $\langle F\rangle$ or $\langle R \rangle$. It is a function of $u$ and the two angular degrees of freedom encoded in the null vector $\hat{x}^\mu = (1,{\hat{\bf x}})$.}  {Inserting complete sets of final states into the expectation value, and using the mode expansion of $F_{\mu\nu}$ and $R_{\mu\nu\sigma\rho}$, one easily obtains an expression for the waveform in terms of scattering amplitudes on the background.}
{The leading contribution is at 1PB, meaning order $e$ (the fundamental charge) in QED or order $\kappa$ (the gravitational coupling) in gravity, but all orders in the background fields}, and comes from interference between \emph{tree-level} 2-point and 3-point  amplitudes.
Unlike in vacuum, 2-point amplitudes on backgrounds are not trivial even at tree-level, encoding e.g.~memory effects~\cite{Adamo:2022rmp}.  {Defining the (theory-dependent) combination
\be
    \alpha(k) = \int\!\ud \Phi(p') \bra{\Psi}{\cal S}^\dagger \ket{p'}\bra{p',k^{\eta}}{\cal S} \ket{\Psi}
\ee
we arrive at, in QED and gravity respectively,
\begin{align}\label{eq:W-G}
    W_{\mu\nu}(u,{\hat x}) &= -\frac{\hbar^\frac{1}{2}}{{\pi}}\text{Re}\int\limits_0^\infty\!\hat{\ud}{\omega}\, \e^{-i\omega u}\, k_{ [\mu} \: \varepsilon^{-\eta}_{\nu]} \alpha(k) \;, \\
\nonumber
    W_{\mu\nu\sigma\rho}(u,{\hat x}) &= -\frac{\kappa}{\pi \hbar^\frac{1}{2}}\text{Im}
    \int\limits_0^\infty\! \hat{\ud}\omega \, \e^{-i\omega u}\, k_{ [\mu}\varepsilon^{-\eta}_{\nu]} k_{[\sigma}\varepsilon^{-\eta}_{\rho]} \alpha(k)\;,
\end{align}
in which $k_\mu = \hbar \omega {\hat x}_\mu$} for $\omega$ a classical frequency (as will be useful later when taking the classical limit), $\varepsilon_\mu^\eta \equiv \varepsilon_\mu^\eta(k)$ is the photon polarisation vector {and $\hat{\ud}x:=\ud x/(2\pi)$.}
{One can check that the combination of amplitudes in $\alpha(k)$ reproduces} the radiation emitted due to geodesic motion, i.e.~the first contribution of self-force effects~\cite{Poisson:2011nh}. 

\paragraph{Plane wave backgrounds.} Plane waves are highly symmetric vacuum solutions of the Einstein or Maxwell equations with two functional degrees of freedom. In gravity, they are described by metrics of the form~\cite{Brinkmann:1925fr}:
\be\label{eq:brink-metric}
    \ud s^2= 2 \ud x^{\LCp} \ud x^{\LCm} - \ud x^a \ud x^a - \kappa\,H_{ab}(x^{\LCm}) x^a x^b\, (\ud x^{\LCm})^2  \,,
\ee
where Latin indices label the `transverse' directions $x^\LCperp=(x^1,x^2)$, while the $2\times2$ matrix $H_{ab}(x^{\LCm})$ is symmetric, traceless and compactly supported on $x^{\LCm}_i < x^{\LCm} < x^{\LCm}_f$ (ensuring the spacetime admits an S-matrix~\cite{Gibbons:1975jb}). The metric has a covariantly constant null Killing vector $n = \partial_{\LCp}$ (or $n_{\mu} = \delta^{\LCm}_{\mu}$) which will recur throughout. To ease notation, we absorb the gravitational coupling into the background, taking $\kappa H_{ab}\to H_{ab}$ from here on; as such, note that  expressions below containing all orders in $H$ implicitly contain all-order PM contributions in $\kappa$.

Plane wave metrics have several associated geometric structures. First, there is a zweibein $E^a_i (x^-)$ and its inverse $E^{i\,a}(x^-)$, labelled by the index $i = 1, 2$ satisfying  {$\ddot{E}_{i\,a} = H_{ab} E^{b}_{i}$, $\dot{E}_{[i}^a\, E^{\phantom{a}}_{j] a} = 0$\;}.
 {The zweibein} encodes gravitational (velocity) memory through the difference
\begin{equation}
    \Delta E^i_a = E^i_a (x^{\LCm} > x^{\LCm}_f) - E^i_a (x^{\LCm}< x^{\LCm}_i) \;,
\end{equation}
which compares the relative transverse positions of two neighbouring geodesics.   {The zweibein also defines a transverse metric $\gamma_{ij}(x^{\LCm}) := E^a_{(i} \, E^{\phantom{a}}_{j) \, a}$ and deformation tensor $\sigma_{ab}(x^{\LCm}) := \dot{E}^i_a \, E^{\phantom{a}}_{i \, b}$}, the latter encoding the expansion and shear of the null geodesic congruence associated to (\ref{eq:brink-metric}).  These definitions are completed by the initial condition
$E^i_a (x^{\LCm} < x^{\LCm}_i ) = \delta^i_a$, which yields $\gamma_{ij} (x^{\LCm} < x^{\LCm}_i ) = \delta_{ij}$ and $\sigma_{ab}(x^{\LCm} < x^{\LCm}_i ) = 0$.

Turning to electromagnetism, plane waves can be defined by the potential $A_{\mu}(x) = - x^b\, {\mathrm E}_b (x^{\LCm})\, n_{\mu}$ in lightfront coordinates (given by the flat space part of (\ref{eq:brink-metric})) and $n_{\mu}$ is as above.
${\mathrm E}_b(x^{\LCm})$ is the two-component, compactly supported electric field. A useful associated quantity is
\begin{equation}
    a_\LCperp (x^{\LCm} ) :=
    \int_{- \infty}^{x^{\LCm}} \ud s \, {\mathrm E}_\LCperp (s) \;,
\end{equation}
{such that $e a_\LCperp$ is the effective `work done' on a charge}. The electromagnetic velocity memory effect is encoded in the constant {$e a_\LCperp(x^\LCm>x^\LCm_f)$}~\cite{Dinu:2012tj}; this is the change in transverse momentum of a particle crossing the background from the asymptotic past to the future.

To simplify the presentation of our results we make the assumption that velocity memory effects induced by our backgrounds are parametrically small, and thus negligible. (We relax this assumption in Appendix~\ref{app:Impulse}.)
This means setting $a_{b}(x^\LCm>x^\LCm_f)=0$ in electromagnetism, and $E^i_a(x^\LCm>x^\LCm_f)=\delta^i_a$ in gravity. {The main simplification is that the tree-level 2-point amplitudes reduce to  {$\bra{p'}{\cal S}\ket{\Psi} \to \e^{i\theta(p')}\phi(p')$, for a theory dependent phase} $\theta$ which can be absorbed by}  {redefining $u$~\footnote{In general the phase will however encode position memory effects on the scattered scalar~\cite{Ilderton:2013dba}.}}.
%
\section{Electromagnetism}\label{sec:QED}
 {We now construct} the classical limit of the electromagnetic waveform $W_{\mu\nu}(u,\hat{x})$ from (\ref{eq:W-G}). Given our assumption of no memory, the only ingredient required is the  {3-point amplitude for a charged scalar, on an electromagnetic plane wave background, to emit a photon. Let the incoming (outgoing) scalar have momentum $p_{\mu}$ ($p'_{\mu}$), and the emitted photon have momentum $k_\mu$ and helicity $\eta$.} The amplitude is calculated by evaluating the cubic part of the action on the appropriate scattering states in a plane wave, see e.g.~\cite{Fedotov:2022ely}. The result is
\begin{align}
\nonumber
&\langle p', k^{\eta}| \mathcal{S}| \Psi \rangle =
    \int\!\ud \Phi(p)\phi(p)\, \e^{i p \cdot b /\hbar}\,
    \hat{\delta}^{3}_{\LCp,\LCperp} (p'+k-p)
    \mathcal{A}_{3} \;, \\
    \label{eq:genform}
    &\mathcal{A}_{3}
    =
    -\frac{2 i e}{\hbar^{3/2}} \int_y \, \varepsilon^{\eta} \cdot P (y) \exp\bigg[{\displaystyle\frac{i}{\hbar} \int_{-\infty}^{y} \ud z\,\frac{k\cdot P(z)}{p_\LCp - k_\LCp}}\bigg],
\end{align}
{where $\int_y:=\int_{-\infty}^{\infty}\! \ud y$ and $\hat{\delta}(x):=2\pi\delta(x)$.} The `dressed' momentum $P_\mu(y)$ is the classical momentum of the particle in the background,
\begin{equation}\label{eq:EM-dressed}
    P_{\mu}(y) = p_{\mu} - e a_{\mu}(y) + n_{\mu} \frac{2 e a(y) \cdot p - e^2a^2 (y)}{2p_+} \;,
\end{equation}
where $a_{\mu}(y)=\delta_{\mu}^{\perp}a_{\perp}(y)$, obeying $P^2(y)=m^2$. Only three components of overall momentum are conserved  {in $\mathcal{A}_3$} as the background breaks  {$x^\LCm$-translation} symmetry.


\paragraph{Calculation of the waveform.} We assemble the QED waveform in (\ref{eq:W-G}) from (\ref{eq:genform}), {using the assumption of negligible memory effects.}
 {We perform the sum over photon helicities} using the completeness relation in lightfront gauge. All gauge-dependent pieces vanish by anti-symmetry or generate boundary terms which can be ignored~\cite{Dinu:2012tj}, leaving only a contribution from $-\eta_{\mu\nu}$. 
An immediate simplification  {in the classical limit} is that the delta function sets $p'=p$, and thus the wavepacket appears as $|\phi(p)|^2$. This means that the impact parameter $b$ drops out, and under the usual assumption that $\phi$ is sharply peaked around some classical momentum, we can integrate over $p$, localising the integrand at the on-shell momentum of the incoming particle, which we continue to write as $p$ for simplicity. This gives
\be\label{eq:W-QED-near-final}
    W_{\mu\nu}(u,{\hat x}) =
    -\frac{i e}{ 4\pi^2 p_\LCp}
    \int_{y,\omega}\!\!\! {\omega}\,
    \e^{-i{ \omega}(u- {\hat x}\cdot X(y))}\,
    {{\hat x}_{[\mu} P_{\nu]}(y)} \;,
\ee
in which $X^\mu(y)$ is the classical particle orbit, obeying~$X'_\mu(y)= P_\mu(y)/p_\LCp$.  {Performing the frequency integral yields a very compact final expression for the classical waveform:}
\be\begin{split}
    W_{\mu\nu}(u,{\hat x}) &=
    \frac{e}{ 2\pi}\int_y \delta(u-{\hat x}\cdot X(y))\, \frac{\ud}{\ud y} \frac{{\hat x}_{[\mu} P_{\nu]}(y)}{{\hat x}\cdot P(y)} \\
    &=\frac{e}{ 2\pi} \sum_\text{sols} \frac{p_\LCp}{{\hat x}\cdot P}\,\frac{\ud}{\ud x} \frac{{\hat x}_{[\mu} P_{\nu]}}{{\hat x}\cdot P} \;,
\end{split}
\ee
where the sum runs over all solutions of the delta-function constraint. It can be checked that this matches the result obtained directly from classical electrodynamics; see Appendix~\ref{app:classical}. 


\paragraph{Properties of the waveform.}
\begin{figure}[t!]
    \centering
    \includegraphics[width=0.7\columnwidth]{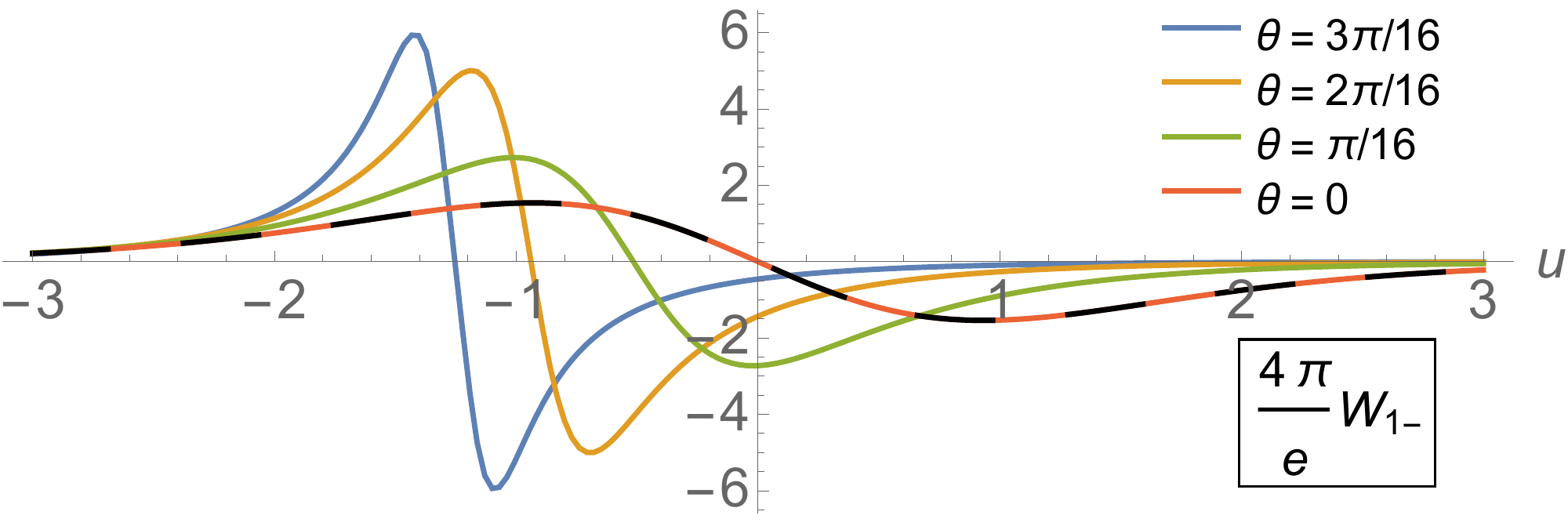}
    \quad
    \includegraphics[width=0.7\columnwidth]{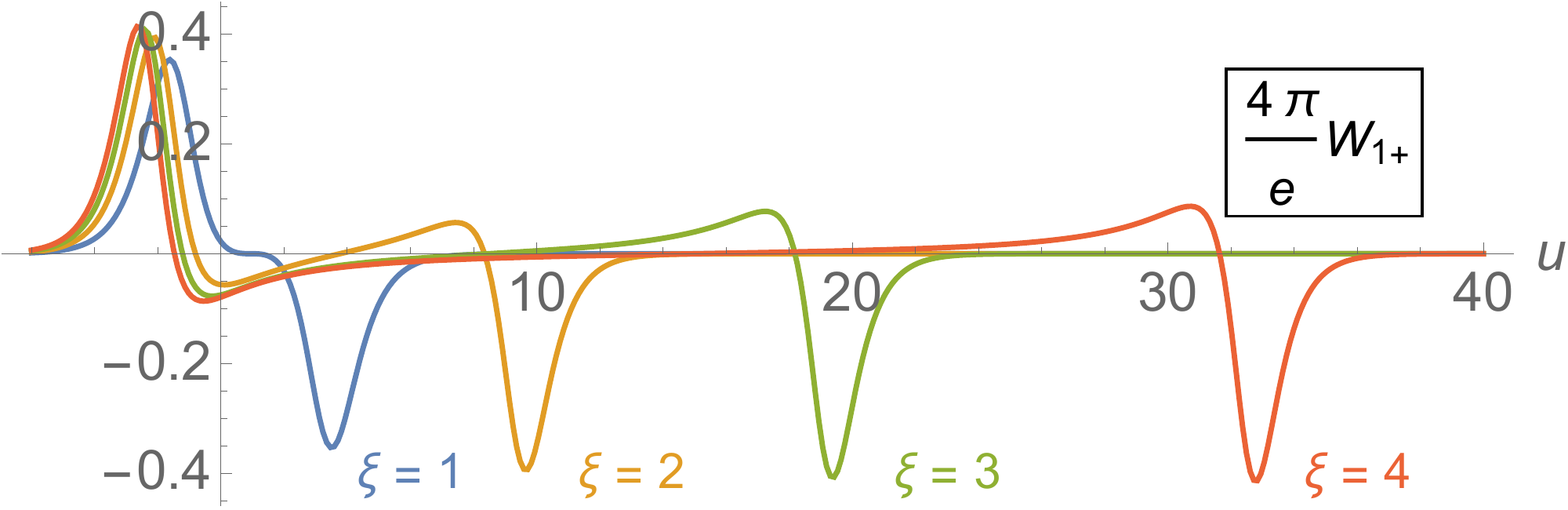}
    \caption{    \label{fig:QED-waveform}
 Two examples of the waveform $W_{\mu\nu}(u,{\hat x})$ for a particle at rest struck by the wave $ea_1=m\xi \text{sech}^2(\nu x^\LCm)$ and $a_2=0$, for strength $\xi$ and frequency $\nu$. We work in units where $\nu=1$. \emph{Upper}: $W_{1\LCm}(u,{\hat x})$ as a function of $u$ for various $\theta$. We have fixed $\xi=2$ and $\phi=0$. At $\theta=0$ (red/black dashed curve), the waveform is a multiple of the driving field $F_{\mu\nu}$ as in (\ref{eq:forward-QED}), but is very different for larger angles. \emph{Lower}: $W_{1\LCp}(u,{\hat x})$ at fixed angles $\theta=\pi$, $\phi=0$, showing  {the dependence of the waveform on the strength $\xi$ of the background.}}
 \end{figure}
First observe that, due to the derivative, the waveform is vanishing in the absence of acceleration. Indeed the final integration by parts, performed as part of the evaluation of the frequency integral, corresponds to removing Coulomb field contributions from the asymptotic waveform, i.e.~restricting to the radiation field which is of interest~\cite{Boca:2009zz}.

Next,  {observe} from \eqref{eq:EM-dressed} that the dressed momentum $P$, hence the orbit $X$, is quadratic in the  {coupling $e$}: it follows immediately that the waveform contains terms of \emph{all} orders in~$e$. This is both explicit, due to the presence of $P$ in the denominator, and implicit, in that one must solve the delta-function constraint. This requires inverting ${\hat x}\cdot X(y)$ which will introduce arbitrary \emph{non-polynomial} dependence on the coupling. (Even for  {the simple but unphysical choice of a} `box' electric field, solving the constraint means solving a cubic equation.) In general, there will be \emph{multiple} solutions to the constraint, meaning that the waveform at any given $(u,{\hat x}_\mu)$ is sourced at several points on  {the orbit}. 

We examine $W_{\mu\nu}$ by choosing a specific plane wave profile and other kinematic data; Fig.~\ref{fig:QED-waveform} illustrates the rich structure found in the classical waveform for a `Sauter pulse' defined by $ea_1=m\xi \text{sech}^2(\nu y^\LCm)$ and $a_2=0$ for strength $\xi$ and frequency $\nu$. Furthermore, the all-orders property of the waveform can be made explicit in the case of an impulsive plane wave, for which all integrals can be performed; see Appendix~\ref{app:Impulse}.  Alternatively, we can expand in powers of $e$, recovering the first perturbative contribution to our waveform, coming from Compton scattering in vacuum~\cite{Cristofoli:2021vyo}; see Appendix~\ref{app:comparison}.

For any plane wave, we can consider the waveform aligned with the direction of the background: ${\hat x}_\mu = \sqrt{2}n_\mu$ (the factor results from conventions). Parameterising ${\hat x}^\mu$ by azimuthal and polar angles $\phi$ and $\theta$, respectively, alignment with the background corresponds to $\theta=0$. At this collinear point the argument of the delta function is simply $u-\sqrt{2} y$, and thus has a single point of support. Most of the structure in the waveform vanishes due to contraction or commutation with $n_\mu$, and one finds 
\be\label{eq:forward-QED}
    W_{\mu\nu}\big|_{\theta = 0} = -\frac{e^2}{4\pi} \frac{F_{\mu\nu}\big(\tfrac{u}{\sqrt{2}}\big)}{p_+\sqrt{2}} \;,
\ee
a result we will later contrast with gravity. If we consider any other point on the celestial sphere, the waveform has a far richer structure, though -- see again Fig.~\ref{fig:QED-waveform}.

\section{Gravity}\label{sec:gravity}
%
We now require the tree-level 3-point amplitude for a massive scalar emitting a graviton, on the gravitational plane wave background. Let the on-shell incoming/outgoing momentum for the scalar be  $p_{\mu}$/$p'_{\mu}$, but let $k_\mu$ now be the emitted graviton momentum. In contrast to QED, all particles are `dressed' in gravity: in scattering calculations, any particle of asymptotic momentum $l_{\mu}$ and mass $m$ has the dressed momentum~\cite{Adamo:2017nia,Adamo:2020qru}
\begin{align}
\label{dressed-momentum}
L_{\mu}(y)\ud y^{\mu} &= l_\LCp \ud y^\LCp + \big( l_i E^i_a + l_+ \sigma_{ab} y^b \big) \ud y^a \\
    &+  \Big( \frac{m^2}{2l_+} + \gamma^{ij}\frac{l_i l_j}{2l_\LCp} + \frac{l_\LCp}{2} \dot{\sigma}_{bc} y^b y^c + l_i \dot{E}_b^i y^b\Big) \ud y^\LCm \,, \nonumber
\end{align}
which obeys $g^{\mu\nu}L_{\mu}(y)L_{\nu}(y)=m^2$. Note that, in contrast to the dressed momentum (\ref{eq:EM-dressed}) in QED, the gravitational dressing depends on the perpendicular coordinates $y^{a}$.  The {outgoing} graviton polarisation also becomes `dressed' by the background; it is conveniently expressed in terms of a projector acting on the free polarisation: 
\begin{align}
\label{eq:grav-polarisation}
    &{{\mathcal{E}}}_{\mu \nu}^{\eta} (k;y) = 
    \mathbb{P}_{\mu\nu\sigma\rho}(k;y){\varepsilon}^{\sigma \rho}_{\eta} \\
    &:=   \bigg[ \mathbb{P}_{\mu \rho}(k; y) \mathbb{P}_{\nu \sigma}(k; y) - \frac{{i \hbar}}{k_\LCp} n_{\mu} n_{\nu} \delta^a_{\rho} \delta^b_{\sigma} \sigma_{ab} (y) \bigg]{\varepsilon}^{\sigma \rho}_{\eta} \;,  \nonumber
\end{align}
where $\mathbb{P}_{\mu \nu}(k;y) = g_{\mu \nu}(y) -2 K_{(\mu }(y)n_{\nu)}/k_\LCp$ contains the dressed momentum $K_\mu(y)$ of the graviton. With these ingredients and the simplification of negligible memory, we can write down the required amplitude~\cite{Adamo:2020qru}:
\begin{equation}
    \begin{gathered}\label{eq:three-point-grav}
        {\mathcal{A}_3} = -\frac{{2}i \kappa}{{\hbar^{3/2}}} \int_y \frac{\exp \big[ i \mathcal{V}(y)\big]}{\sqrt{|E(y)|}}\,
       {{ {\cal E}}^{\eta}_{\mu\nu}(k ; y) P^{\mu}(y) P'^{\nu}(y)}\,,        
    \end{gathered}
\end{equation}
 {where the first line of} (\ref{eq:genform}) still holds, the exponent is
\begin{equation}\label{eq:V-thing}
    \mathcal{V}(y) := {\frac{1}{\hbar}}\int_{-\infty}^{y^-} \ud z\, \frac{P_{\mu}(z) K_{\nu}(z) g^{\mu \nu}(z)}{p_+ - k_+}\,,
\end{equation}
and $|E(y)|$ is the zweibein determinant. It can be checked that all contractions between dressed momenta and polarisations appearing are independent of the transverse coordinates, even though their constituents are not. Hence the integrand in (\ref{eq:three-point-grav}) is a function of only $y^\LCm$, and is (trivially) evaluated on the classical particle orbit parametrized by~$y^\LCm$.

{The calculation proceeds as in QED; we assemble the waveform (\ref{eq:W-G}) from the three-point amplitude (\ref{eq:three-point-grav}).
Similarly to the QED} case, we can restrict the sum over graviton polarisations to physical degrees of  {freedom}.
To obtain the classical limit of the waveform, we inspect powers of $\hbar$ in the amplitude (\ref{eq:three-point-grav}) and in the definition (\ref{eq:W-G});
we again find that all pre-factors of $\hbar$ cancel, and the classical limit is obtained by setting $\hbar=0$ everywhere else. This again allows the wavepacket to be integrated out, arriving at
\be\label{eq:W-G-classical1}
\begin{split}
    &W_{\mu\nu\sigma\rho}(u,{\hat x}) = \frac{\kappa^2}{\pi } \int\limits_0^\infty\!{\hat \ud}\omega\, { \omega}^2\, \e^{-i\omega u}\, {\hat x}_{[\mu}{\hat x}_{[\sigma}
    \int_{y} \frac{\e^{i \omega {\bar {\cal V}}(y)}}{\sqrt{|E(y)|}}\, \\
   &
    \times\,\Big[
    \eta_{\nu]\gamma}\eta_{\rho]\delta} -\frac{1}{2}\eta_{\nu]\rho]}\eta_{\gamma\delta}
    \Big]
    {\bar {\mathbb{P}}}^{\alpha\beta\gamma\delta}(\hat{x},y)P_{\alpha}(y)P_{\beta}(y)
    + \text{c.c.} \;,
\end{split}
\ee
in which the reduced exponent ${\bar{\cal V}}$ is
\begin{align}
    \frac{1}{2p_\LCp}\int^{y}_{-\infty} &\mathrm{d}z\: \frac{m^2}{p_\LCp} {\hat x}^\LCp + \gamma^{i j}(z)\bigg(\frac{p_\LCp}{{\hat x}_\LCp} {\hat x}_{i} {\hat x}_{j}+\frac{{\hat x}_\LCp}{p_\LCp} p_{i} p_{j} -2p_{i} {\hat x}_{j}\bigg) \nonumber\\
    &= {\hat x}\cdot X(y)\,, \label{eq:t-mu-omega}
\end{align}
for $X(y)$ the classical particle orbit and $\bar{\mathbb{P}}_{\mu\nu\sigma\rho}(\hat{x},y) := \mathbb{P}_{\mu\nu\sigma\rho}(k,y)|_{{\bar k}={\omega}{\hat x}}$ evaluated on that orbit. 
Now, the trace-like term in \eqref{eq:W-G-classical1} arising from the polarisation sum can be simplified by first observing that
\be\nonumber
    \eta_{\gamma\delta}\, {\bar {\mathbb{P}}}^{\alpha\beta\gamma\delta}(\hat{x},y)P_{\alpha}(y)P_{\beta}(y) = m^2 +\frac{2ip_\LCp^2}{\omega\hat{x}_\LCp}\bigg[ i \partial_\LCm {\bar{\cal V}} - \frac{1}{2}\sigma^a_{a}\bigg](y) \;.
\ee
{It can be} checked that the term in brackets is exactly the $y^\LCm$ derivative of the entire integrand in (\ref{eq:W-G-classical1}), and hence gives a boundary term which can be dropped, leaving only the mass term.

It remains to perform the $\omega$ integral. However, in contrast to QED, the projector $\bar {\mathbb{P}}^{\alpha\beta\gamma\delta}(\hat{x},y)$ contains terms with different scaling in ${\omega}$. We highlight this by defining 
\begin{align}
T^0_{\nu\rho}(\hat{x},y) &:=\frac{{\mathbb P}_{\nu\alpha}(\hat{x},y){\mathbb P}_{\rho\beta}(\hat{x},y)P^\alpha(y) P^\beta(y) - \frac12 \eta_{\nu\rho} m^2}{\sqrt{|E(y)|}} \;, \nonumber \\
    {T}^{1}_{\nu\rho}(\hat{x},y) &:= \frac{\delta_\nu^a \delta_\rho^b \sigma_{ab}(y)}{{\hat x}_\LCp \sqrt{|E(y)|}}p_\LCp^2 \;, \label{eq:eff-EM-tensors}
\end{align}
such that the integrand scales in the frequency as $\sim {\omega}^2 T^0 - i {\omega} T^{1}$. Combining the presented term in \eqref{eq:W-G-classical1} with its complex conjugate and trading explicit ${\omega}$ factors for $y^{-}$--derivatives gives our final result for the waveform:
\be\label{eq:W-G-classical-final}
\begin{split}
    W_{\mu\nu\sigma\rho}(u,{\hat x}) = -\frac{\kappa^2}{\pi }   {\hat x}_{ [\mu}{\hat x}_{[\sigma}
    \int_{y} \delta(u - {\bar{\cal V}}(y))\,\, \\
    \Big[\mathcal{D}^2\, {T}^0_{\rho]\nu]}(\hat{x},y) - \mathcal{D}\, {T}^{1}_{\rho]\nu]}(\hat{x},y)\Big]\,,
\end{split}
\ee
in which the derivative $\mathcal{D}$ acts as
\be\label{final-D-def}
    \mathcal{D}f(y) := \frac{\ud}{\ud y}\left( \frac{f(y)}{\partial_\LCm {\bar{\cal V}(y)}}\right)\,.
\ee
Again, for confirmation of this result via classical general relativity calculations, see Appendix~\ref{app:classical}.

\paragraph{Properties of the waveform.}\label{sec:small-angle}
%
Some insight into the gravitational waveform is provided by observing from \eqref{eq:t-mu-omega} that $\bar{\mathcal{V}}$ is determined by the 0PB classical orbit $X^{\mu}(y)$ of a particle crossing the plane wave spacetime. The orbit itself goes like the integral of the transverse metric $\gamma^{ij}=E^{(i|a|}E^{j)}_{a}$. Reinstating explicit dependence on the gravitational coupling by taking $H_{ab}\to\kappa H_{ab}$, it is clear that the integral of $\gamma^{ij}$ will contain terms which are at least linear in $\kappa$. Since \eqref{eq:W-G-classical-final} contains terms which go like $\bar{\mathcal{V}}^{-1}$, as well as an integral localised in terms of $\bar{\mathcal{V}}$, it follows that the waveform will contain terms of \emph{all orders} in the background and hence in $\kappa$. To connect to the PM construction of the waveform we expand in $\kappa$, showing in Appendix~\ref{app:comparison} that the leading contribution comes from gravitational Compton scattering.

While the non-linearity of general relativity makes it harder to evaluate the waveform analytically for test plane wave profiles, progress can be made in the impulsive case where $\kappa H_{ab}(x^-)=\delta(x^-)\,\kappa\,\mathrm{diag}(\lambda,-\lambda)$. This is demonstrated in Appendix~\ref{app:Impulse}: the resulting waveform is explicitly all-orders in $\kappa\lambda$. See also~\cite{Ilderton:2018lsf}.

The structure of \eqref{eq:W-G-classical-final} indicates the presence of \emph{tail} effects in the gravitational waveform. This follows from the fact that the two terms in the waveform descend directly from those in the polarization tensor~\eqref{eq:grav-polarisation}. The background dressing of this polarization is directly related to the failure of Huygens' principle for gravitational perturbations in plane wave spacetimes: initial data localized on a lightcone spreads outside of the lightcone as it evolves~\cite{Friedlander:2010eqa,Harte:2013dba,Adamo:2017nia}. These effects are present in both the $T^{0}$ and $T^1$ terms of the 1PB waveform, with the $T^1$ contribution being pure tail; by comparison, in the PM expansion of the two-body problem tail effects only emerge at fourth-order (e.g., \cite{Bern:2021yeh}).

These tail effects are a consequence of the inherent non-linearity of gravity compared to electromagnetism, and this leads to another interesting feature of the gravitational waveform which is not present in QED. Consider the case, as in (\ref{eq:forward-QED}), where the direction of observation ${\hat x}^\mu$ aligns with the wave direction $n^\mu$, corresponding to azimuthal angle $\theta=0$. 
 {The plane wave metric is not asymptotically flat in precisely this (and only this) direction~\cite{Penrose:1965rx}}, so we approach it with caution. For any $\theta\not=0$ the gravitational waveform is well-defined, but in the limit $\theta\to 0$, it is divergent. To see this, one expands ${\hat x}_\mu$ for small-$\theta$, i.e.~${\hat x}_j = \sin \theta \{\cos\phi,\sin\phi\} \sim \theta$, and 
\begin{equation*}
\begin{split}
    {\hat x}_\LCp = \frac{1-\cos\theta}{\sqrt{2}} \sim \theta^2 \;, \quad
    {\hat x}_\LCm = \frac{1+\cos\theta}{\sqrt{2}} \sim 1 \;.
\end{split}
\end{equation*}
With this, it is simplest to pick components of $W$, and to focus on the pure tail term which contains the deformation tensor $\sigma$. The contribution of this term to $W_{-a-b}$~is
\be\begin{split}
    \frac{\kappa^2 p_\LCp^2 {\hat x}_\LCm {\hat x}_\LCm}{\pi {\hat x}_\LCp}  
    \int_{y} \delta(u - {\bar{\cal V}}(y))\,\,
    \mathcal{D}\, \frac{\sigma_{ab}(y)}{\sqrt{|E(y)|}}  {\sim \frac{1}{\theta^2}} \;.
\end{split}
\ee
in which the $1/{\hat x}_\LCp$ term generates the divergence ( {while}
${\bar{\cal V}}$ and $\partial_\LCm {\bar{\cal V}}$ remain finite in the limit $\theta\to 0$). The divergence reflects the fact that it is not possible to `scatter' gravitons in the $n_\mu$ direction, in which the background is not asymptotically flat; the interaction between the emitted radiation and the background never switches off. {This is in contrast to QED, where the photon and background do not interact, and the waveform remains finite, c.f.~(\ref{eq:forward-QED}). (Indeed the distinction with QED is visible at the entirely perturbative level of the scalar-graviton Compton amplitude, which is singular at forward scattering~\cite{Bjerrum-Bohr:2014lea}.)}  The angular divergence would have physical consequences; it will enter, via the Riemann tensor, into the geodesic deviation equation for a null congruence at the next order of the PB expansion. The divergence will thus  emerge as a physical singularity describing a region of spacetime in which null geodesics become infinitely separated. It would be interesting to investigate this.

%
\section{Conclusions}\label{sec:concs}
We have derived the gravitational waveform emitted by a massive particle when it scatters off a gravitational plane wave background, a solution to the fully non linear Einstein equations. Analogous formulae have been presented for the electromagnetic case. In contrast to existing results, these waveforms are manifestly \emph{all-orders} in the coupling, and exhibit a rich structure including tail effects that usually enter at higher order in the PM expansion. Our results underline the power of using strong field amplitudes to study classical physics~\cite{Adamo:2022rmp}. In future work we aim to go to higher orders in the PB expansion, including higher points and loops. There is no conceptual obstacle to doing so, and we expect this to provide easier access to observables of interest in classical gravity. It would also be interesting to consider other physically relevant strong backgrounds, like black holes or beams of gravitational radiation, and to analyse our results for specific profiles arising as plane wave limits of these backgrounds.


\acknowledgments

We thank Tom Heinzl for interesting conversations. The authors are supported by a Royal Society University Research Fellowship (TA), Leverhulme Trust grant RPG-2020-386 (TA \& AC), and an EPSRC studentship~(SK).

\appendix

\section{Classical checks}\label{app:classical}
This appendix contains a classical derivation of the waveforms in electromagnetism and gravity. Schematically, these stem from radiation fields `$A$' generated by sources `$J$' representing particles moving on a background, which take the form
\begin{equation}
    A_{\sigma}(x) := \int \ud^4 y \, G_{\mathrm{ret}} (x,y) \,J_{\sigma}(y) \label{eq:green-funct-integral} \;,
\end{equation}
in which the subscript ${\sigma}$ is a placeholder for any number of vector indices or spin labels. The retarded Green's function is the inverse of $\nabla^2$ in a flat or curved background, and is therefore theory-dependent:
\begin{align}
    &G^{\mathrm{EM}}_{\mathrm{ret}} (x-y) = i \Theta (x^- - y^-) \int \frac{\hat{\ud}\bar{k}_{+}\hat{\ud}^2\bar{k}_{\perp}}{2\bar{k}_+} \Theta(\bar{k}_{+})\: \e^{-i\bar{k} \cdot (x - y) } \;,\label{eq:QED-greens-func}\\
    &G^{\mathrm{GR}}_{\mathrm{ret}} (x, y) = \frac{i\Theta (x^\LCm - y^\LCm)}{\sqrt{|E(x)|}}  \int \frac{\hat{\ud}\bar{k}_{+}\hat{\ud}^2\bar{k}_{\perp}}{2\bar{k}_+} \Theta(\bar{k}_{+}) \frac{\e^{-i \mathcal{F}(x, y)} }{\sqrt{|E(y)|}}  \;, \label{eq:GR-greens-func}
\end{align}
in which $\bar{k}_\mu$ is on-shell and 
\be
    \mathcal{F}(x, y) = \bar{k}_\LCp (x - y)^\LCp + \bar{k}_i (E_a^i (x) x^a - E_a^i (y) y^a) + \frac{\bar{k}_\LCp}{2} (\sigma_{ab}(x) x^a x^b - \sigma_{ab}(y) y^a y^b ) + \frac{\bar{k}_i \bar{k}_j}{2\bar{k}_\LCp} \int_{y^\LCm}^{x^\LCm} \ud s \, \gamma^{ij}(s).
\ee
We measure the waveform at future null infinity, hence we write the coordinate $x^\mu$ in the coordinate system $(u, r, \mathbf{\hat{x}})$ where $r = |\mathbf{x}|$, $\mathbf{x} = r \mathbf{\hat{x}}$ and $u = t- r$; the asymptotic limit is reached by taking $r \rightarrow \infty$ at fixed $u$ and angular coordinates $\mathbf{\hat{x}}$. As long as our measurement device is not in the beam of the wave (corresponding to $\hat{x}^3 = 1$), then we can set the initial $\sqrt{|E(x)|} = 1$ in \eqref{eq:GR-greens-func} to unity. The step function can also be set to unity in the limit. With this, the `Fourier transformed' version of \eqref{eq:green-funct-integral} is
\begin{equation}
    A_{\sigma}(x) = i\int \frac{\hat{\ud}\bar{k}_{+}\hat{\ud}^2\bar{k}_{\perp}}{2\bar{k}_+}  \e^{-i \bar{k} \cdot x} \, \bar{J}_{\sigma}(\bar{k}) \;,
\end{equation}
where $\bar{J}_{\sigma}(\bar{k})$ theory-dependent. In the $r \rightarrow \infty$ limit the leading behaviour of this quantity is, performing a saddle point calculation of the $\bar{k}_\LCperp$ integrals as in the text,
\begin{equation}
    A_{\sigma} (x) \sim  \frac{1}{4 \pi r} \int_0^{\infty} \hat{\ud}\omega \:\e^{- i \omega u} \bar{J}_{\sigma} (\omega \hat{x}_{\mu} ) \,+ \, \text{c.c.}\,, \label{eq:asymptotic-waveform-class}
\end{equation}
where $\hat{x}_{\mu} = (1, \mathbf{\hat{x}})$ in Cartesian coordinates. We now turn to specifics in electromagnetism and gravity.

\subsubsection*{Electromagnetism}
In electromagnetism $J$ is the vector current for a particle moving in a background field,
\begin{equation}
    \bar{J}_{\mu}(\bar{k}) = - e\int_{y} \e^{i \bar{k} \cdot X} \frac{\partial}{\partial y} \frac{X'_{\mu}(y)}{i {\bar k} \cdot X'(y)}
\end{equation}
where $X_{\mu}(y)$ is the particle orbit, and dashes represent derivatives with respect to $y^\LCm$. Note that this form of the current generates the radiation field; the Coulomb fields from outside the wave have been subtracted. Substituting into \eqref{eq:asymptotic-waveform-class} we obtain the gauge potential of the radiation field:
\begin{equation}
      A_{\mu} (u, \hat{x} ) \sim  \frac{ie}{4 \pi r} \int_0^{\infty} \hat{\ud}\omega \int_{y}  \,  \e^{-i \omega ( u - \hat{x} \cdot X (y)} \frac{1}{\omega} \frac{\partial}{\partial y} \frac{X'_{\mu}(y)}{\hat{x} \cdot X'(y)} \,+\, \mathrm{c.c.} \label{eq:this}
\end{equation}
The radiated field strength is $F_{\mu \nu} = 2 \partial_{[\mu} A_{\nu]}$ which, up to subleading corrections in $1/r$, we can obtain by adding factors of $-i\omega \hat{x}$ to the integrand of (\ref{eq:this}). The factors of $\omega$ outside the exponential cancel in $F_{\mu\nu}$ allowing us to perform the $\omega$-integral to find
\begin{equation}
    F_{\mu \nu} (u, \hat{x}) \sim \frac{1}{r} W_{\mu\nu}(u, \hat{x} )=\frac{e}{2 \pi r} \int_{y} \, \delta( u - \hat{x} \cdot X(y) )\frac{\partial}{\partial y} \frac{\hat{x}^{\phantom{a}}_{[\mu}X'_{\nu]}(y)}{\hat{x} \cdot X'(y)} \;,
\end{equation}
where `$\sim$' denotes equality up to subleading terms in $r^{-1}$. This exactly the waveform derived in the text from the classical limit of the quantum result.

\subsubsection*{Gravity}
%
The gravitational radiation of a massive scalar moving in a background is sourced by the stress-energy tensor
\begin{equation}
    T_{\mu \nu}(y) =  \frac{P_{\mu}(y) P_{\nu}(y)}{p_+}\, \delta^{3}_{+, \perp}(y - X(y))\,,
\end{equation}
in which $X_{\mu}(y)$ is once again the particle orbit and $P_{\mu}(y) = p_\LCp X'_{\mu}(y)$. For convenience we define $\widetilde{T}_{\mu \nu} = T_{\mu \nu} - \frac{1}{2} g_{\mu \nu} T^{\alpha}_{\alpha}$ as shorthand for a `trace-reversed' $T_{\mu\nu}$. From the Einstein field equations, one can derive that the sourced gravitational field satisfies, imposing lightfront gauge $n^{\mu}h_{\mu\nu} = 0$,
\begin{equation}
    h_{\mu \nu} = - \frac{2}{\nabla^2} t_{\mu \nu} + \frac{2}{\nabla^2} n_{\mu} n_{\nu} h^{ab} H_{ab} \;. \label{eq:class-grav-eq}
\end{equation}
in which the modified stress-energy tensor $t_{\mu \nu}$ is defined by
\begin{equation}
    {t}_{\mu \nu} := \kappa^2 \Bigg[ {\widetilde T}_{\mu \nu} - 2 \frac{\nabla_{(\mu} \widetilde{T}_{\nu)+}}{\partial_+} + \nabla_{(\mu} \nabla_{\nu)} \frac{{\widetilde{T}}_{++}}{\partial_+^2}\Bigg] \;. \label{eq:modified-EM-tensor}
\end{equation}
In the notation of  \eqref{eq:green-funct-integral}, the source `$J$' is now
\be
    \bar{t}_{\mu \nu} (k) = 2 \kappa^2 \int_{x_i^-}^{x_f^-} \ud y \: \e^{i \mathcal{V}(y)}\,\frac{\partial}{\partial y} \Bigg[ \frac{1}{\sqrt{|E(y)|}} \Bigg( \mathbb{P}_{\mu \alpha} \mathbb{P}_{\nu \beta} P^{\alpha} P^{\beta} - \frac{1}{2} \eta_{\mu \nu} m^2 
    - \frac{ip_+^2 \delta_{\mu}^a \delta_{\nu}^b \sigma_{ab}}{k_+} \Bigg)\frac{1}{i \partial_- \mathcal{V} (y) }\Bigg]\,,
\ee
where $\mathbb{P}_{\mu \nu}(k; y^-)$ are the projectors defined in \eqref{eq:grav-polarisation} and
\begin{equation}
    \mathcal{V}(y) = \int_{-\infty}^{y} \ud z\, \frac{g^{\mu\nu}(z)\, \bar{K}_{\mu}(z) \, P_{\nu}(z)}{p_+}\,.
\end{equation}
To arrive at this expression one uses integration by parts to shift the derivatives present in \eqref{eq:modified-EM-tensor} onto the propagator. Additionally, we ignore the second term in \eqref{eq:class-grav-eq} since we are only interested in radiative contributions. See~\cite{Adamo:2020qru} for details. Substituting into \eqref{eq:asymptotic-waveform-class} we obtain an expression for the asymptotic metric perturbation
\be
   h_{\mu \nu} (u, \hat{x}) =  \frac{-i \kappa^2 }{2 \pi r} \int_0^{\infty} \hat{\ud}\omega \int_{y_i^-}^{y_f^-} \!\ud y \, \frac{\e^{- i \omega (u - \bar{\mathcal{V}}(y))}}{\omega} 
   \frac{\partial}{\partial y} \Bigg[ \Big(T^0_{\mu \nu} - \frac{i}{\omega}T^{1}_{\mu \nu}\Big) \frac{1}{\partial_- \bar{\mathcal{V}} (y)}\Bigg] + \mathrm{c.c}.\,,
\ee
in which the same `effective' energy-momentum tensors defined in~\eqref{eq:eff-EM-tensors} have appeared, along with with the reduced exponent
\begin{equation}
    \bar{\mathcal{V}}(y) = \frac{1}{\omega} \mathcal{V}(y)
    \Big|_{\bar{k} = \omega \hat{x}}.
\end{equation}
From here we form the linearised curvature $R_{\mu \nu \sigma \rho} = 2 \partial_{[\mu} \partial_{[\sigma} h_{\nu] \rho]}$, with each derivative introducing a factor of $(-i)\omega \hat{x}$ into the integrand. We can then integrate in $\omega$ to obtain
\begin{equation}
    R_{\mu \nu \sigma \rho}(u, \hat{x})\sim - \frac{\kappa^2}{\pi r} \hat{x}_{[\mu} \hat{x}_{[\sigma} \int_{y} \delta (u - \bar{\mathcal{V}}(y^\LCm) ) \Big[ \mathcal{D}^2 T^0_{\nu] \rho]}(u, \hat{x}) - \mathcal{D} T^{1}_{\nu] \rho]}(u, \hat{x}) \Big] \;,
\end{equation}
where $\mathcal{D}$ is defined in (\ref{final-D-def}). This confirms the classical limit of our QFT calculations.

\section{The impulsive case}\label{app:Impulse}
%
In this appendix we calculate the waveform for impulsive plane wave backgrounds.
%
\subsubsection*{Electrodynamics}
%
An impulsive plane has electric fields $\mathrm{E}_\perp(x^-) = \mathrm{E}_\perp \delta(x^\LCm)$, for $\mathrm{E}_\perp$ constant. We write $a_\mu = \delta_\mu^\perp \mathrm{E}_\perp\,\Theta(x^\LCm) \equiv \mathrm{E}_\mu\, \Theta(x^\LCm)$. An incoming particle with momentum $p_\mu$ for $x^\LCm<0$ is kicked by the wave to momentum $P_\mu$ at $x^\LCm>0$ where
\be\label{P-inf}
    P_\mu = p_\mu -e \mathrm{E}_\mu + n_\mu\, \frac{2e\mathrm{E}\cdot p - e^2\,\mathrm{E}\cdot \mathrm{E}}{2n\cdot p} \;.
\ee
This is a memory effect, which we neglected in the text. However, the addition of memory does not impact the final expression for the electrodynamics waveform, which holds for \emph{any} plane wave.  The waveform is most easily evaluated using (\ref{eq:W-QED-near-final}) by splitting the $\ud x^\LCm$ integral into two parts: $x^\LCm \gtrless 0$. The remainder of the calculation is trivial, and one finds
\be\label{eq:W-QWED-imp}
    W_{\mu\nu}(u,{\hat x})  = \frac{e}{2\pi}\, \delta(u) \bigg[ \frac{{\hat x}_{[\mu} P_{\nu]}}{{\hat x}\cdot P} -\frac{{\hat x}_{[\mu}\, p_{\nu]}}{{\hat x}\cdot p}  \bigg] \;.
\ee
The waveform manifestly contains terms all orders in the coupling $e$. It is supported on the same singularity structure in $u$ as the driving electric field is in $x^\LCm$. The tensor structure clearly derives directly from standard soft factors for momentum transfer $p\to P$. Neglecting memory in this case amounts to assuming $\mathrm{E}_\mu$ is parametrically small, in which case one replaces $P\to p$ and the waveform vanishes. For an impulsive background, the waveform is thus `pure memory.'

\subsubsection*{Gravity}
%
The situation is gravity is rather more intricate, even for an impulsive background, and the structures provide an interesting contrast with electrodynamics. The impulsive metric is given by taking, in (\ref{eq:brink-metric}), $\kappa H_{ab}(x^\LCm)\to \kappa \delta(x^\LCm) H_{ab}$ with $H_{ab}$ now constant (though still symmetric and traceless). We can, without loss of generality, choose coordinates to diagonalise $H_{ab}=\mathrm{diag}(\lambda,-\lambda)$. In contrast to QED, the memory effects present in this metric cannot, in general, be directly treated with the expressions in the text. In order to present the full impulsive result, we compute the relevant amplitudes directly, using the wavefunctions in~\cite{Adamo:2017nia}.

The momentum kick is given by replacing {$e \mathrm{E}_\mu \to \kappa p_\LCp \delta^a_\mu H_{ac}b^c$} in (\ref{P-inf}), in which $b$ is the transverse impact parameter (a dependence not present in electrodynamics). From here there are, as in electrodynamics, two contributions: one from before the impulse ($x^-<0$) and one from after the impulse ($x^->0$). That from  $x^\LCm>0$ yields a term similar to (\ref{eq:W-QWED-imp}), while that from $x^\LCm<0$ is more complicated. One finds
\begin{align}\label{shebang1}
    W_{\mu\nu\sigma\rho}(u,{\hat x}) &=
        \frac{\kappa^2}{4\pi}\,
        \delta'(u) \, {\hat x}_{ [\mu} {\hat x}_{[\sigma}  \varepsilon^{-\eta}_{\rho]}
        \varepsilon^{-\eta}_{\nu]}
        {\varepsilon}^\eta_{\alpha}
        {\varepsilon}^\eta_{\beta}
        \bigg[\frac{P^\alpha P^\beta}{{\hat x}\cdot P} - \frac{p^\alpha p^\beta}{{\hat x}\cdot p}\bigg] \\
    &+\frac{i\kappa^2}{4\pi}
    {\hat x}_{ [\mu} {\hat x}_{[\sigma}
        \varepsilon^{-\eta}_{\rho]}
        \varepsilon^{-\eta}_{\nu]}
    \int_0^\infty\! {\hat \ud}\omega\, \omega^2 e^{-i\omega u}\!\int\!{\hat \d}^2\ell_\LCperp\, (\varepsilon^{\eta}(\ell)\cdot p)^2
    \frac{1}{\ell\cdot p}  \mathcal{F}(\ell-k) +  \, \text{c.c.,} \nonumber
\end{align}
in which $k_\mu \equiv \omega {\hat x}_\mu$, $\varepsilon_\mu$ \emph{without} argument is $\varepsilon_\mu({\hat x})$,
$\ell_\mu$ is a null vector with fixed longitudinal component $\ell_\LCp = k_\LCp$, and
\be
    {\cal F}(\ell) := \frac{\e^{ i b^a \ell_a \, -i\ell^a H_{ab}^{-1} \ell^b/(2k_\LCp)}}{{\kappa\lambda k_\LCp}} - {\hat \delta}_\LCperp(\ell) \;.
\ee
Consider the first line in (\ref{shebang1}); the tensor structure is a double copy of the soft factor structure in electrodynamics, but the singularity is now $\delta'(u)$ rather than $\delta(u)$. The second line of (\ref{shebang1}) is, in a sense, `pure tail' and we have not yet found a very compact expression for the remaining integrals for general ${\hat x}^\mu$ and $u$. {Nevertheless, the result as presented is clearly of all orders in $\kappa$.}

Moreover, we can consider a special case which allows a direct, if tedious, calculation of \emph{all} terms in the impulsive waveform. First, we choose $p_\LCperp=0$, so that the wave-particle collision is `head on'. Second, we choose the impact parameter as $b^\LCperp=0$; this has the effect of turning off memory, since the kicked momentum $P$ collapses back to incoming $p$. Finally, we choose a particular point of observation on the celestial sphere, $\theta=\pi$ (antipodal to the direction in which the background is not asymptotically flat), which sets ${\hat x}_\mu \to \sqrt{2} \delta_\mu^\LCp$. In this case, the first line of (\ref{shebang1}) vanishes -- this is clearly due to the assumption of no memory, as in QED. The second line remains and simplifies considerably. By performing the angular integration in $\ell_\LCperp$, one finds that the remaining integrals are independent of helicity and the helicity sum can then be performed directly.

From here one writes the $\omega$ factors as derivatives with respect to $u$ and combines the presented term in (\ref{shebang1}) with its complex conjugate. This gives, writing $q \equiv |\ell_\LCperp|$,
\be
    W_{\mu\nu\sigma\rho} = \kappa^2 \frac{p_\LCp^3}{\kappa\,\lambda}\, \delta^\LCp_{[\mu} \delta^\LCp_{[\sigma}  (-1)^{(a)} \delta_{\rho]}^a \delta_{\nu]}^a \partial_u^2 \int\limits_0^\infty {\hat \ud} q \, \frac{{\hat \ud} q\,q^3}{2m^2+p_\LCp^2 q^2} \int\limits_0^\infty{\hat \ud}\omega\, J_1\bigg(\frac{\omega\,q^2}{\kappa\,\lambda\sqrt{8}}\bigg) \cos (\omega u) \;,
\ee
in which a sum over $a\in(1,2)$ is implied and $J_1$ is a Bessel function of the first kind. The remaining integrals may be evaluated using~\cite{Gradshteyn:1702455}, resulting in
\be\label{shebang2}
    W_{\mu\nu\sigma\rho} = -\frac{\kappa^2\,p_\LCp}{\pi^2\sqrt{8}} \,  \delta^\LCp_{[\mu} \delta^\LCp_{[\sigma}  (-1)^{(a)} \delta_{\rho]}^a \delta_{\nu]}^a\,\, \frac{\partial^2}{\partial u^2} \left(\frac{\nu \log (\nu + \sqrt{\nu^2-1})}{\sqrt{\nu^2-1}}\right) \,,
\ee
\begin{equation*}
\mbox{for } \quad \nu := \kappa\,\lambda\,\sqrt{2}\frac{p_\LCp^2}{m^2}\, |u| \;.
\end{equation*}
Thus, unlike the `soft' terms in the first line of \eqref{shebang1}, the `pure tail' terms are not localised. Taking the derivatives in (\ref{shebang2}), one finds that they do include a localised piece at the origin, proportional to $\delta(u)$ like in electrodynamics, rather than $\delta'(u)$ as in the first line of (\ref{shebang1}). This is not unexpected if the waveform is supported entirely on $T^1$: recalling the discussion around (\ref{eq:eff-EM-tensors}), in Fourier space the contribution from $T^1$ carries the same frequency dependence as electrodynamics.

\section{Comparison with perturbative amplitudes}\label{app:comparison}

Here, we establish that the weak-field limit of our calculations reproduces well-known, standard approaches to classical physics from perturbative scattering amplitudes~\cite{Kosower:2018adc,Cristofoli:2021vyo}. We treat the gravitational case explicitly, with the electromagnetic calculation following similar lines.

We begin by reinstating all factors of $\kappa$, with the aim of expanding the 3-point amplitude \eqref{eq:three-point-grav} in powers of $\kappa$ to make contact with the PM expansion around flat spacetime. Note the overall, explicit factor of $\kappa$ in \eqref{eq:three-point-grav}: this means that the plane wave background can be completely switched off by simply setting $\kappa=0$ everywhere else in the expression. What remains in this case is the scalar $\to$ scalar + graviton 3-point amplitude in Minkowski spacetime, which vanishes on the support of momentum conservation. Thus, the first non-trivial contribution to the perturbative expansion of \eqref{eq:three-point-grav} is at order $\kappa^2$, as expected for a \emph{four}-point tree-level amplitude in flat spacetime.

To make this explicit, we begin by identifying the leading order perturbative expression for all quantities associated with the plane wave metric. This begins with the zweibein $E^a_i (x^-)$ obeying  {$\ddot{E}_{i\,a} = \kappa H_{ab} E^{b}_{i}$. Writing
    $E^i_a = \delta^i_a + \mathcal{O}(\kappa)$, in which the first term imposes the boundary conditions, we immediately find
    \[
        E_{ia}(x^\LCm) = \delta_{ia} + \kappa  I_{ia}(x^\LCm) \quad \text{where} \quad I_{ab}''(x^\LCm) = H_{ab}(x^\LCm) \;.
    \] 
It is easily checked that the Wronskian condition $\dot{E}_{[i}^a\, E^{\phantom{a}}_{j] a} = 0$ is obeyed at order $\kappa$, and the remaining geometric objects associated to the wave follow as, also to order $\kappa$,
    \[
        \gamma_{ij}(x^\LCm) = \delta_{ij} + 2 \kappa I_{ij}(x^\LCm)\,,
        \qquad
        \gamma^{ij}(x^\LCm) = \delta^{ij} - 2 \kappa I^{ij}(x^\LCm) \;,
        \qquad
        \sigma_{ab}(x^\LCm) = -\kappa I'_{ab}(x^\LCm) \;,
        \qquad
        |E|= 1 \;.
    \]
Observe that to linear order in $\kappa$, all indices are raised and lowered with the flat metric.

This is enough to specify the dressed momentum (\ref{dressed-momentum}), the projectors in and below (\ref{eq:grav-polarisation}), and the exponent $\mathcal{V}$ in (\ref{eq:V-thing}). As such, all factors appearing in the integrand of the amplitude \eqref{eq:three-point-grav} can be expanded, with only the overall linear contribution in $\kappa$ being retained. This calculation is lengthy, but direct. 

Some simplifications occur upon using the explicit form of the lightfront-gauge polarisation vectors (obeying $n\cdot\varepsilon(k)=k\cdot\varepsilon(k)=0$) and taking the graviton polarization to be the trace-free symmetric product of these spin-1 polarizations (i.e., $\varepsilon_{\mu\nu}=\varepsilon_{(\mu}\varepsilon_{\nu)}$). The contraction between these polarizations and an arbitrary 4-vector $v^\mu$ gives
    \be\label{eq:trikset}
        \varepsilon_\eta(k)\cdot v = \epsilon^\eta_a \left( \frac{v_\LCp}{k_\LCp}\, k_a -v_a\right) \;,
    \ee
    in which $\epsilon^\eta_a =(1, i \eta)/\sqrt{2}$ are transverse vectors carrying the helicity label $\eta=\pm1$. With this, one finds the weak-field expansion of the gravitational three-point amplitude:
\be
\begin{split}
    {\mathcal{A}_3} \simeq i \kappa^2&\,
	\hat{\delta}^3_{\LCp,\LCperp}(p'+k-p)\, \frac{p_\LCp}{k_\LCp}\,  (p_\LCp-k_\LCp)  \\
 &\int\!\ud x^\LCm\, \e^{i \frac{k\cdot p\,x^-}{p_\LCp-k_\LCp}}\, \frac{\left[ (\varepsilon_{\eta}(k) \cdot p) S^a - (k \cdot p) \varepsilon^a_{\eta}(k) \right]}{k \cdot p}\, I'_{a b}(x^\LCm)\, \frac{\left[ (\varepsilon_{\eta}(k) \cdot p) S^b - (k \cdot p) \varepsilon^b_{\eta}(k) \right]}{k \cdot p} \,, 
\end{split}
\ee
in which ``$\simeq$'' denotes equality at order $\kappa^2$, $S_a := k_a - p_a (k_\LCp/p_\LCp)$ and we have used integration by parts in $x^\LCm$ to bring the amplitude to this compact `product' form, the importance of which will become clear in a moment. 

The integral over $x^\LCm$, which has been present from the start, can now be performed due to the perturbative limit. It yields the Fourier transform of $I'_{ab}$, hence ~$-i\omega I(\omega)_{ab}$, evaluated at $\omega := k\cdot p/(p_\LCp-k_\LCp)$. The next step is to understand how $I_{ab}$ relates to the frequency and polarisation of the gravitons which make up the wave. To do so, we write the plane wave metric in self-dual and anti-self dual parts. It is equivalent, and more convenient, to make this decomposition for $I_{ab}$, though:
\[
    I_{ab}\equiv
    \begin{pmatrix}
    I_{11} & I_{12} \\ I_{12} & -I_{11}
    \end{pmatrix}
     \to \frac12
    \begin{pmatrix}
    f_1 + f_{-1} & i(f_1 - f_{-1})\\
    i(f_1 - f_{-1}) & -f_1- f_{-1}
    \end{pmatrix} \;,
\]
in which $f_{1}$ and $f_{-1}=f_1^\star$ are the self-dual and anti-self dual parts respectively. This choice of notation follows from the observation that $I_{ab}$ can be written in terms of the helicity 2-vectors introduced above as
\[  I_{ab}(\omega)
=\sum_{s= \pm 1} f_s(\omega)\, \epsilon^s_a\, \epsilon^s_b \;.
\]
This enables the perturbative limit to be written as a sum over helicities:
\be
   {\mathcal{A}_3} \simeq  \kappa^2
	\hat{\delta}^3_{\LCp,\LCperp}(p'+k-p) 
 \sum_{s=\pm}f_s(\omega)\, \frac{p_\LCp}{k_\LCp}\,  (p_\LCp-k_\LCp)\,  \omega\, \bigg(\frac{(\varepsilon_{\eta}(k) \cdot p) S^a - (k \cdot p) \varepsilon^a_{\eta}(k)}{p\cdot k}\, \epsilon^s_{a}\bigg)^2 \,, 
\ee
Writing $\ell_\mu=\omega n_\mu$ for the momentum of the incoming graviton resulting from the perturbative limit of the plane wave background, we then observe that $\varepsilon^s(\ell)=(0,0,1,is)/\sqrt{2}$ are a corresponding basis of helicity vectors in the chosen lightfront gauge. Using (\ref{eq:trikset}), and the conservation of momentum implied by the three delta functions, we arrive at:
\be
{\mathcal{A}_3} \simeq 
    \kappa^2
	\hat{\delta}^3_{\LCp,\LCperp}(p'+k-p) 
 \sum_{s=\pm}f_s(\omega)\, 
 \frac{\ell\cdot p\, \ell\cdot p'}{k\cdot \ell}\, \bigg( \frac{\varepsilon_s \cdot p'\, \varepsilon_\eta\cdot p}{p\cdot \ell} -\frac{\varepsilon_s \cdot p\, \varepsilon_\eta\cdot p'}{p\cdot k}  - \varepsilon_\eta\cdot \varepsilon_s \bigg)^2   \;,
\ee
for the perturbative amplitude.

Up to overall constants which can be absorbed into the profiles $f_s$, the squared term in large brackets is precisely the scalar QED Compton amplitude. The factor outside the brackets completes the known double-copy prescription for transforming that amplitude into, as promised, the tree-level gravitational Compton amplitude stripped of overall momentum-conserving delta functions, see~\cite{Choi:1994ax,Bern:2002kj,Bjerrum-Bohr:2014lea, Holstein:2017dwn}. Even momentum conservation is obtained from the perturbative limit, though: to consider the scattering of a single graviton rather than a wave, we must have $f_\sigma(\omega) \sim \delta(\omega - \omega_\star)$ where $\omega_\star$ is the chosen initial frequency. This gives a fourth delta functions which, it is easily checked, combines with the other three to give precisely the expected $\hat{\delta}^4(p'+k - p - \ell)$ for Compton scattering. Note that the perturbative limit of the 3-point amplitude for scalar QED in a plane wave follows similar lines (cf., \cite{Adamo:2020qru}).

Having established that the leading contribution to the weak-field limit of the 3-point amplitude on a plane wave amplitude is the gravitational Compton amplitude in Minkowski spacetime, it follows that the weak-field limit of our gravitational waveform agrees with the leading-order waveform in the standard PM expansion. To see this, one simply observes that the leading PM contribution to the classical radiation produced by scattering a scalar probe with a coherent state of gravitational radiation is controlled by the tree-level Compton amplitude involving the incoming and outgoing scalars, the emitted graviton, and an incoming graviton (taken from the coherent state)~\cite{Cristofoli:2021vyo}. Thus, if the perturbative limit of our strong-field amplitude produces this Compton amplitude, agreement with the leading PM waveform follows automatically.

Expanding the strong-field amplitudes to higher-orders in $\kappa$ will likewise lead to higher-order perturbative results, although it should be stressed that these will only correspond to a \emph{portion} of the full PM answer at each order. This is because every perturbative photon/graviton extracted from the plane wave background will have its momentum lying along $n_{\mu}$, so the resulting amplitudes will be for multicollinear configurations in a trivial vacuum (cf., \cite{Adamo:2021hno}, and see~\cite{Seipt:2013hda,HernandezAcosta:2019vok,Golub:2020kkc} for applications in QED).


\bibliographystyle{JHEP}
\bibliography{NewBib}

\end{document}